\documentclass[twocolumn]{revtex4}
\usepackage{amsmath,amssymb,amsthm,bm,dcolumn,pst-all,pstcol}
\usepackage[dvips]{graphicx}
\usepackage[latin1]{inputenc}
\newcommand{\ket}[1]{\ensuremath{\left|#1\right\rangle}}
\newcommand{\bra}[1]{\ensuremath{\left\langle#1\right|}}

\newcommand{\parcial}[2]{\ensuremath{\frac{\partial#1}{\partial #2}}}

\newcommand{\deri}[2]{\ensuremath{\frac{d #1}{d #2}}}

\def\A{\mathbf{A}}
\def\r{\mathbf{r}}
\def\p{\mathbf{p}}
\def\e{\mathbf{e}}
\begin{document}
%
\title{Some aspects about gauge transformations in non-relativistic QED}
\author{J. A. Sanchez-Monroy$^{1,}$}
\email{jasanchezm@unal.edu.co}
\author{J. Morales$^{1,2}$}
\email{jmoralesa@unal.edu.co}
\affiliation{Grupo de Campos y Part\'\i culas$^1$,}
\affiliation{Centro Internacional de F\'\i sica$^2$\\Bogot\'a, Colombia}
\author{D. E. Zambrano}
\email{dezambranor@unal.edu.co}
\affiliation{Centro Brasileiro de Pesquisas F\'\i sicas - CBPF, Rio de Janeiro Brazil}
\begin{abstract}
The wave mechanical formulation of quantum electrodynamics is investigated under gauge transformations. For this purpose we observe the structure of Schr\"odinger equation and matricial elements under these transformations. 
We conclude this theory is not gauge invariant since the eiger-energies may depend of gauge function, since that the Hamiltonian is operator gauge dependent and cannot represent the observable energy.
\end{abstract}
\maketitle


\section{Introduction}
Is a well-known fact that physical laws cannot depend of arbitrarities arisen from mathematical formalism.
There are some cases where arbitrarities are introduced to make easy the calculations, but these quantities cannot be  experimentally measured \cite{st1,momento}.
\newline
\par
\textit{Gauge transformations} in electrodynamics are an example of mathematical arbitrarity. Classically this theory is gauge invariant \cite{Ja9,ed}. A. M. Stewart studied the effect of gauge transformations over Hamiltonian and Schr\"odinger equation with classical potentials (\emph{semiclassical electrodynamics} regime \cite{st1,momento}), he found that only one class of gauge functions can leave the results invariant \cite{st1}. There are a stronger restriction for conservative systems \cite{st2}, which is perhaps due to some troubles in the interpretation of the Hamiltonian as energy \cite{dfg, cohen}. 
\newline
\par
In this work we show that semiclassical electrodynamics is not gauge invariant\cite{st1,st2}, and we obtian the restriction for gauge functions to preserve the invariance\cite{momento} ({\it see} section II). We prove in section III that although there exists a restriction, which relates potentials that produce the same physical results, there not exist a way to determine an unique solution for a physical system (At the Appendix we propose a possible theorerical alternavite). Finally, in the section V we dicuss the results.
\section{Gauge dependence of semiclassical theory}
Let $A$ and $\phi$ be the electrodynamical potentials that describes a particular electromagnetic field.
We may apply the \emph{gauge} transformation
\begin{equation}
\label{tg}
\A\to\A'=\A+\nabla\chi,\quad\quad
\phi\to\phi'=\phi-\parcial{\chi}{t}
\end{equation}
where $\chi$ is a scalar function called {\it gauge function}. If the equations of motion do not change, the dynamics of the system is conserved and the theory is invariant under gauge transformations. Classical electrodynamics is gauge invariant\cite{Ja9,momento}. 
\par
In quantum mechanics with the Schr\"odinger picture, Schr\"odinger equations gives the temporal evolution of the system,
\begin{equation}\label{Schrodinger}
H\Psi(\r,t)=i\hbar\frac{\partial}{\partial t}\Psi(\r,t).
\end{equation}
Hamiltonian operator under a gauge transformation is 
\begin{equation}
\label{Hx}
H_{\chi}({\bf p}_{\chi},{\bf r},t)=\frac{\left({\bf
p}_{\chi}-e({\bf
A}+\nabla\chi)\right)^2}{2m}+e\left(\phi-\frac{\partial\chi}{\partial
t}\right),
\end{equation}
where subindex $\chi$ denotes the gauge function, it is clear that this gauge transformation does not change the picture. For $\chi= 0$,  
\begin{equation}
H_0=\frac1{2m}\left(\p-\frac{e}{c}\A\right)^2+e\phi .
\end{equation}

We assume that exist an unitary operator $U(\chi)$ such that $\Psi'(\r,t)=U\Psi_0(\r,t)$. In general $U$ is not time-independent, since $\chi$ may depend explicitly of time. Multiplying to the left the Schr\"odinger equation for $\Psi_0$ by $U$, we obtain that
\begin{equation}\label{SchrPsi}
i\hbar U\parcial{\Psi_0}{t}=UH_0\Psi_0.
\end{equation}
The chain rule for derivatives provides us the identity
\begin{equation}
U\parcial{\Psi_0}{t}=\parcial{(U\Psi_0)}{t}-\parcial{U}{t}\Psi_0.
\end{equation}
Replacing the last expression in Eq. \eqref{SchrPsi} we obtain that
\begin{equation}
\begin{split}
i\hbar\parcial{(U\Psi_0)}{t}&=UH_0\Psi_0+i\hbar\parcial{U}{t}\Psi_0\\&=\left(UH_0U^{\dag}+i\hbar\parcial{U}{t}U^{\dag}\right)(U\Psi_0).
\end{split}
\end{equation}
The relation between $\Psi_0$ and $\Psi'$ allows us to write that
\begin{equation}
i\hbar\parcial{\Psi'}{t}=H'\Psi'=\left(UH_0U^{\dag}+i\hbar\parcial{U}{t}U^{\dag}\right)\Psi',
\end{equation}
then the form of the Schr\"odinger equation's remains invariant under any transformation with the form
\begin{equation}
H'=UHU^{\dag}+i\hbar\parcial{U}{t}U^{\dag}.
\end{equation}
Taking $U(\chi)={\bf e}^{ie\chi(\r,t)/h}$, we obtain the Hamiltonian under a gauge transformation Eq. \eqref{Hx}. Explicitly, $\Psi_0$ transforms as
\begin{equation}\label{Psi->}
\Psi_0
\to\Psi_{\chi}
=U(\chi)\Psi_0({\bf
r},t)=\Psi_0({\bf r},t){\bf e}^{ie\chi({\bf r},t)/\hbar}.
\end{equation} 
Here is important to emphasise that wavefunctions and Hamiltonian change locally with a gauge transformation, then 
Schr\"odinger equation may change and thus the physics of the system. In order to see this we calculate the matricial elements of the Hamiltonian operator. Taking $\ket{n_0}$ and $\ket{m_0}$, which are two elements of the basis of Hamiltonian $H_0$, the matricial element after a gauge transformation is given by  
\begin{equation}
\begin{split}
\label{mx|H|nx=m|H|n} 
\bra{n_{\chi}}H_{\chi}\ket{m_{\chi}}
&
=\bra{n_0}H_0\ket{m_0}-e\bra{n_0}\parcial{\chi}{t}\ket{m_0}\\
&
=E_{n,0}\delta_{n,m}-e\int\parcial{\chi}{t}\psi^*_{n,0}(\r,t)\psi_{m,0}(\r,t)\,d\r
\end{split}
\end{equation}
where $\ket{n_{\chi}}:=\e^{ie\chi/\hbar}\psi_{n,0}(\r,t)$ and $E_{n,0}$ is the eigenvalue related to $\ket{n_0}$. Hence the Hamiltonian might be not diagonal in the transformed basis. 
\par
In the other hand, the experimental measures correspond to energy differences, which must be invariant. From Eq. \eqref{mx|H|nx=m|H|n} we can see that 
\begin{multline}\label{11}
E_{n,\chi}-E_{m,\chi}=E_{n,0}-E_{m,0}\\+e\int\parcial{\chi}{t}\left[|\rho_{m,0}(\r,t)|^2-|\rho_{n,0}(\r,t)|^2\right]\,d\r
\end{multline} 
where $\rho_{n,0}(\r,t)$ is the probability density related to $\ket{n_0}$. In conclusion, energy differences eventually are not invariant. If we restrict to $\chi$ by means
\begin{equation}
\label{forma}
\chi(\r,t)=f(\r)+g(t),
\end{equation}
we guarantee the invariance of energy differences.
\par
We have another trouble when we consider a time-independent Hamiltonian, which we transform by means a gauge function as in Eq. \eqref{forma}. Eq. (\ref{mx|H|nx=m|H|n}) implies that
\begin{equation}\label{13} 
\begin{split}
E_{n,\chi}&
=E_{n,0}-e\int\psi*_{n,0}(\r,t)\parcial{\chi}{t}\psi_{n,0}(\r,t)\,d\r\\
&
=E_{n,0}-e\deri{g(t)}{t},
\end{split}
\end{equation}
so in order to keep eigenvalues time-independent of a time-independent Hamiltonian under a gauge transformation, the gauge function should fulfills 
$$
\parcial{\chi}{t}=const.
$$
This equation implies that 
\begin{equation}\label{la que es}
\chi(\r,t)=f(\r)+kt
\end{equation} 
with $k=const$. Eq. \eqref{la que es} is the restriction for gauge functions that guarantee the invariance of the theory.
\section{Deep inside trouble}
We define $\mathfrak{U}$ as set of all electrodynamical potentials possible that \emph{describe} a given system.  Eq. \eqref{la que es} divides the set $\mathfrak{U}$ in equivalence classes. Each equivalence class has associated a family of Hamiltonians, which produce the same physical results
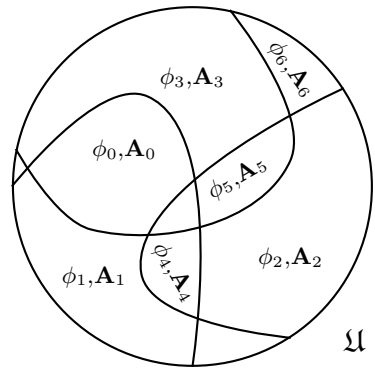
\begin{figure}
\centering
\psset{unit=1.2cm}
\begin{pspicture}(-2,-2)(2,2)
\pscircle(0,0){2}
\pscurve(0,-2)(-.5,1)(-2,0)
\rput{258}(0,0){\pscurve(0,-2)(.7,-1)(-.5,1)(-2,0)}
\rput{123}(0,0){\pscurve(0,-2)(-.5,1)(-2,0)}

\rput(-.7535,.4){$\phi_0$,$\textbf{A}_0$}
\rput(-1.09,-1){$\phi_1$,$\textbf{A}_1$}
\rput(1.09,-.8){$\phi_2$,$\textbf{A}_2$}
\rput(0,1.20){$\phi_{3}$,$\textbf{A}_{3}$}
\rput{-60}(-.2,-.950){\small $\phi_{4}$,$\textbf{A}_{4}$}
\rput{30}(.50,.10){\small $\phi_5$,$\textbf{A}_5$}
\rput{-55}(1.1,1.30){$\phi_{6}$,$\textbf{A}_{6}$}
\rput(1.8,-1.7){\large $\mathfrak{U}$}

\end{pspicture}
\caption{All potentials space $\mathfrak{U}$. 
Each piece of this circle corresponds to a set of potentials that are related between themselves by \eqref{la que es}. For each set we may choice a rep, dividing $\mathfrak{U}$ on equivalence classes (sets of potentials). Only a one of this equivalence class describes the actual behaviour of the system (actual universe).\label{dibujo}} 
\end{figure} 

\par
If we take two different equivalence classes, we obtain two different families of Hamiltonians that yield contradictory results. Since there not exist a physical criterion to choice the correct family, thus we have two different solutions for the same system. Eq. \eqref{tg} allow us to obtain an infinity (non-numerable) number of families of Hamiltonians, that in principle \emph{describe} the same electromagnetic field, but from a mathematical point of view we may obtain infinite number of distinct solutions for the same system (see FIG \ref{dibujo}).
\par
In order to illustrate this situation, we consider the Harmonic Oscillator: a particle inside of an electric field $\mathbb{E}=-mw^2x/e$. Two possible Hamiltonians for this system are 
\begin{eqnarray}
\label{por}
H_0&=&
-\frac{\hbar^2}{2m}\nabla^2+\frac{1}{2} mw^2 x^2,
\\\label{por2}
H_1&=&
-\frac{\hbar^2}{2m}\nabla^2+\frac{1}{2} mw^2 x^2-2ekt,
\end{eqnarray}
where $k$ is a constant that adjusts the units. Is clear that $H_0$ and $H_1$ are related by the gauge transformation $\chi =kt^2$ and thus two both belong to two different families of Hamiltonians. 
Eigenvalues for each Hamiltonian are
\begin{align}
E^{0}_n&=\hbar\omega\left(n+\frac12\right),& n&=0,1,2,\dots,\\
E^{1}_n&=\hbar\omega\left(n+\frac12\right)+2ekt,& n&=0,1,2,\dots,
\end{align}
respectively. The last results are distinct and contradictory, since first one describes a conservative system and the other one does not.
\par
At this point, we cannot decide which family of Hamiltonian are correct, there are two different families of Hamiltonians that yield contradictory results. We obtain that the Hamiltonian is operator gauge dependent and cannot represent the observable energy. If we consider the hamiltoniano of free particle now we see
\begin{equation}
H=-\frac{\hbar^2}{2m}\nabla^2,
\end{equation} 
which represents the energy. Another hamiltoniano like (which one can obtain by means of a gauge transformation)
\begin{equation}
H=-\frac{\hbar^2}{2m}\nabla^2+e\phi(t),
\end{equation} 
it does not represent the energy for the case of a free particle. Motivated by this case in the Appendix we purpose to fix the gauge searching.
\section{Conclusions}
Non-relativistic QED is not gauge invariant since the eiger-energies may depend of gauge function. In order to maintain the same physical results, gauge function must be have the form
\begin{equation}\setcounter{equation}{12}
\chi=f(\r)+kt,
\end{equation} 
\setcounter{equation}{17}
hence, we cannot take arbitrarily the electrodynamical potentials $\A$ and $\phi$. Restriction \eqref{la que es} divides the set of potentials of a physical system in equivalence classes, this implies the existence of infinite number of distinct solutions for the same system. Since there not exists a physics-theoretical criterion to decide which family of Hamiltonians is proper to describe the system, only experimental result can do it.
\par
With this panorama, we should think to find a theoretical procedure such that selects an unique equivalence class and be suitable for any physical system. In the Appendix we purpose to fix the gauge searching for, where we find a method that allows us to choice an unique family of Hamiltonians. 
\subsection*{Appendix}
Fixing the gauge, maybe we could restrict the potentials to an unique equivalence class. We only must to impose to the set of potentials that fulfills the gauge condition that be a subset of an unique equivalence class defined by \eqref{la que es} and this set contains at least a couple of potentials ($\A,\,\phi$) for each physical problem. The last condition means that the gauge does not restrict the potentials at the point that these potentials just only work for particular cases. 
\par
It is easy to prove that Lorentz gauge $$\nabla\cdot\A+\frac1c\frac{\partial\phi}{\partial t}=0$$ 
and time-independent gauge 
$$\frac{\partial \phi}{\partial t}=0$$ 
do not satisfy these condition. But the {\it modified temporal gauge} \mbox{$(\phi=const.)$,} fulfills it, since
\begin{equation}\label{otra1}
\frac{\partial \chi}{\partial t}=const,
\end{equation}
and for any pair of potentials ($\A,\,\phi$) we can build the potentials ($\A',\,\phi'$) by means
\begin{eqnarray}
\phi'&=&0,\\
\A'&=&\A+\int_T^t\!\!\nabla \phi'\,dt',
\end{eqnarray}
where it is clear that $\A'$ and $\phi'$ describe the same electromagnetic field that $\A$ and $\phi$ and belong to the equivalence class of modified temporal gauge.
%
%

\end{document}